\documentclass[twocolumn,aps,prl,amsmath,amssymb,floatfix]{revtex4}

\usepackage{psfrag}
\usepackage[dvips]{graphicx}

\begin{document}
\title{Entangled black holes as ciphers of hidden information}

\author{Samuel L.\ Braunstein}
\affiliation{Computer Science, University of York, York YO10 5DD, UK}
\author{Hans-J\"urgen Sommers}
\affiliation{Theoretische Physik, Universit\"at Duisburg-Essen,
47048 Duisburg, Germany}
\author{Karol \.{Z}yczkowski}
\affiliation{Institute of Physics, Jagiellonian University,
30-059 Krakow, Poland}
\affiliation{Center for Theoretical Physics,
Polish Academy of Science, 02-668 Warszawa, Poland}

\date{\today}

\begin{abstract}
The black-hole information paradox has fueled a fascinating effort to
reconcile the predictions of general relativity and those of quantum
mechanics. Gravitational considerations teach us that black holes must
trap everything that falls into them. Quantum mechanically the mass of
a black hole leaks away as featureless (Hawking) radiation.
However, if Hawking's analysis turned out to be accurate then the
information would be irretrievably lost and a fundamental axiom of
quantum mechanics, that of unitary evolution, would likewise
fail. Here we show that the information about the
matter that collapses to form a black hole becomes encoded into pure
correlations within a tripartite quantum system, the quantum analog
of a one-time pad until very late in the evaporation,
provided we accept the view that the thermodynamic entropy of a black
hole is due to entropy of entanglement.
In this view the black hole entropy is primarily due to trans-event
horizon entanglement between external modes neighboring the black hole
and internal degrees of freedom of the black hole.
\end{abstract}

\maketitle

A powerful tool for studying black hole evaporation as a unitary process
is in terms of random subsystems. The starting point of this approach is
that the evaporative dynamics can be modeled by sampling a random 
subspace from the black hole interior, of dimensionality equaling 
the radiation subsystem. This idea was originally formulated \cite{Page93}
in a model where all the in-falling matter was in a pure state
$|i\rangle$, so 
\begin{equation}
|i\rangle_{\text{int}}\rightarrow (U|i\rangle)_{\text{RB}}. \label{Page}
\end{equation}
Here the initial internal (int) Hilbert space of the black hole may
be thought of as evolving under a random unitary $U$ followed by
the `emission' of radiation into (a now randomly selected) subspace $R$
with the {\it reduced-size\/} interior labeled $B$. A key assumption
of any random matrix calculation is the dimensionality of the space 
on which the random matrices act. For black hole evaporation it was
argued \cite{Page93} that this dimensionality should be well
approximated from the {\it thermodynamic\/} entropy
$S_{\text{BH}}={\cal A}/4$ of a black hole of area ${\cal A}$, 
giving a dimensionality
$\text{dim}(\text{int})= R B= e^{S_{\text{BH}}}$ --- where we reuse
subspace labels for Hilbert space dimensionalities. Therefore,
the black hole interior comprises $n=\log_2[\text{dim}(\text{int})]$
qubits.

Within this original model, `discernable information' is defined as
the deficit from maximal entropy in the radiation subspace. It was
found that this quantity remains almost zero, until half the qubits
of the entire black hole had been radiated, after which the discernable
information rises at the rate of roughly two bits for every qubit
radiated \cite{Page93}. This behavior suggests that first entanglement
is created, followed by dense coding \cite{Bennett92} of {\it classical\/}
information about the initial state. By entangling the state of the
in-fallen matter in this model with some distant reference (ref) subspace,
one can track the flow of quantum information using information
theoretic tools. In this way Eq.~(\ref{Page}) becomes \cite{Hayden07}
\begin{equation}
\frac{1}{\sqrt{K}}\sum_{i=1}^{K}
|i\rangle_{\text{ref}}\otimes|i\rangle_{\text{int}}\rightarrow
\frac{1}{\sqrt{K}}\sum_{i=1}^{K}
|i\rangle_{\text{ref}}\otimes(U|i\rangle)_{\text{RB}}.
\label{HaydenPreskill}
\end{equation}
Here $k=\log_2 K$ is the number of qubits describing the quantum state
of the matter used to form the black hole. That this is tiny in
comparison to the number of qubits comprising the black hole itself,
$k\lll n$, is one of the signature properties of a black
hole \cite{tHooft93}. Using the decoupling theorem \cite{Abey06} (see
Appendix~\ref{decoupling} for details), we may conclude that prior to
$\frac{1}{2}(n-k)-c$ qubits having been radiated, the quantum information
about the in-fallen matter is encoded within the black hole interior
with fidelity at least $1-2^{-c}$; whereas after a further $k + 2c$
qubits have been radiated, the in-fallen matter is encoded within
the radiation subspace with fidelity at least $1-2^{-c}$ (see also
Ref.~\onlinecite{Hayden07} for this latter result). The quantum
information about the in-fallen matter leaves in a narrow `pulse'
at the radiation emission rate.

Our key point of departure from previous work is motivated by the
well-accepted result from many-body quantum theory that the
entanglement of across a boundary will generically scale as the
boundary's area \cite{Eisert09}. It is therefore natural to conjecture
that a black hole's thermodynamic entropy might be primarily due to
entropy of entanglement between modes external to, but in the
neighborhood of the event horizon and modes of the black hole's
interior. Such a conjecture has indeed been made
\cite{tHooft85,Bombelli86,Srednicki93,Hawking01,Brustein06,Emparan06};
it holds naturally for some models of black holes \cite{Hawking01,Brustein06}
and even resolves some difficulties associated with computing entropy at
the microscopic level \cite{Emparan06}. Here we show that this conjecture
leads to a radically different picture of information flow from black
holes.

As with Ref.~\onlinecite{Hayden07} we tag the information about the
matter that collapsed to form the black hole by entanglement with
some distant reference (ref) subsystem. If we assume that there is
{\it no\/} so-called ``bleaching'' mechanism which can strip
away all or part of the information about the in-fallen matter as it
passes the event horizon, then the initial quantum state of the black hole
interior (int) and its surroundings has the unique form \cite{me}
\begin{equation}
\frac{1}{\sqrt{K}}\sum_{i=1}^K |i\rangle_{\text{ref}}\otimes
\sum_{j}\sqrt{p_j}\,(|i\rangle\otimes |j\rangle\oplus 0)_{\text{int}}
\otimes|j\rangle_{\text{ext}}, \tag{3a} \label{entangledI}
\end{equation}
up to overall int-local and ext-local unitaries.
Here $\oplus 0$ means we pad any unused dimensions of the interior
space by zero vectors \cite{me} and
$\rho_{\text{ext}}=\sum_jp_j|j\rangle_{\text{ext}}%
\,{}_{\text{ext}}\!\langle j|$
is the reduced density matrix for the external (ext) neighborhood modes.
Again we take the dimension of the interior space as
$\text{dim}(\text{int})= R B= e^{S_{\text{BH}}}$.

As with earlier work \cite{Page93,Hayden07}, described by Eqs.~(\ref{Page})
and~(\ref{HaydenPreskill}), we apply a random unitary, constrained
by causality and acting solely on the black hole interior,
$|\psi\rangle_{\text{int}}\rightarrow (U|\psi\rangle)_{RB}$ to
randomly sample the radiation subspace ($R$). Eq.~(\ref{entangledI}) 
then becomes 
\begin{equation}
\rightarrow
\frac{1}{\sqrt{K}}\sum_{i=1}^K |i\rangle_{\text{ref}}\otimes
\!\sum_{j}\sqrt{p_j}\,[U(|i\rangle\otimes |j\rangle\oplus 0)]_{RB}
\otimes|j\rangle_{\text{ext}}.
\tag{3b} \label{entangledF}
\setcounter{equation}{3}
\end{equation}
It will be convenient to define
\begin{eqnarray}
x&\equiv& \log_2(RB/K) +\log_2 ({\text{tr}}\; \rho_{\text{ext}}^2),
\end{eqnarray}
which roughly quantifies the number of excess unentangled qubits
within the initial encoding of the black hole in Eq.~(\ref{entangledI}).
Note that $0\le x \le \log_2(RB/K)$.

Again using the distant reference to tag the information (see
Appendix~\ref{decoupling} for details), it is easy to see that for all
but the final $k+\frac{1}{2}x+c$ qubits radiated, the information about
the in-fallen matter is encoded in the combined space of external
neighborhood modes and black hole interior with fidelity at least
$1-2^{-c}$. Similarly, for all but the initial $k+\frac{1}{2}x+c$
qubits radiated, this information is encoded in the combined radiation
and external neighborhood modes with fidelity at least $1-2^{-c}$.
In addition, at all times this information is encoded with unit
fidelity within the joint radiation and interior subspaces.

In other words, between the initial and final $k+\frac{1}{2}x+c$ qubits
radiated, the information about the in-fallen matter is effectively
deleted from each subsystem individually \cite{me,Kretschmann}, instead
being encoded in any two of the three of subsystems. During this time,
the information about the in-fallen matter is to an excellent
approximation encoded within the perfect correlations of a quantum
one-time pad \cite{me,Leung02} consisting of the three subsystems:
the radiation, the external neighborhood modes, and the black hole
interior. This description applies to the entire evaporation period
except for the short encoding and decoding periods (assuming small 
$x$ above). A heuristic picture showing a smooth flow of information 
is given in Appendix~\ref{heuristic}.

We now consider what happens if additional matter is dumped into the 
black hole after its creation. Following Ref.~\onlinecite{Hayden07}, 
we model this process via cascaded random unitaries on the black hole 
interior\,---\,one unitary before each radiated qubit. Within the pure 
state model of Eq.~(\ref{HaydenPreskill}), it was argued \cite{Hayden07} 
that after half of the initial qubits had radiated away, any information 
about matter subsequently falling into the black hole would be ``reflected'' 
immediately at roughly the radiation emission rate \cite{Hayden07}. A
subtle flaw to this argument is due to the omission of the fact that a
black hole's entropy is non-extensive, typically scaling as the square
of the black hole's mass $M^2$: for every $q$ qubits dumped into
a black hole, the entropy increases by $O(qM)\gg q$. Likewise, the
number of unentangled qubits within the black hole will increase by
$O(qM)$. Therefore, within the cascaded unitary pure-state model,
the reflection described in Ref.~\onlinecite{Hayden07} would not begin
immediately, but only after a large delay in time of $O(qM^2)$.
Notwithstanding the delay, the very different {\it behaviors\/} of
the black hole in the first and second halves of its life endows
it with a kind of quasistatic ``hair'' associated with its history
since creation.

By contrast, the one-time-pad description of evaporation [for black 
holes described by the entropy-as-entanglement conjecture of Eq.~(3)] 
paints a very different picture. Instead of the reflection of information 
found in the pure-state model, here, any additional qubits thrown into 
the black hole will immediately begin to be encoded into the tripartite 
correlation structure (assuming negligible $x$). Therefore, just as in 
the uncascaded case, the decoding into the radiation subspace of {\it 
all} the information about all the in-fallen matter will only occur at 
the very end of the evaporation. The non-extensive increase in black 
hole entropy is taken up as entanglement with external neighborhood 
modes so no further delays occur. 
Importantly, entanglement-based black holes really are ``hairless'': 
their {\it behavior\/} does not qualitatively change in time.

As we have seen, in the pure-state model of a black hole the information
about the in-fallen matter leaves in a narrow pulse after half the qubits
have evaporated, whereas for the entangled-state model, this
information appears in the out-going radiation only at the end of
the evaporation. One way to reconcile these two models is if the
pure-state model were run for twice as many qubits, but stopped
just after the information about the in-fallen matter had escaped.
If we accepted the model of a black hole as highly entangled
across its event horizon we could justify this reconciliation as a
rough approximation: In particular, instead of fixing a boundary
at the event horizon we could fix it somewhat further out, say at
$r=3M$. In this case, the dimensionality enclosed would be roughly
the square of that of the black hole interior space itself by including
the contribution from the external neighborhood modes; this would
therefore yield roughly twice as many qubits as the black hole itself
holds. The trans-boundary entanglement at $r=3M$ would be approximately
static over the entire course of evaporation so it could be factored out
as non-dynamical, thus leaving an approximately pure-state description.
Finally, once the original number of qubits had evaporated away (now
half the total for our modified pure-state model) the black hole interior
would be exhausted of Hilbert space and evaporation would cease. This 
suggests that despite the incompatibility between the two models, a
pure-state analysis, if properly set up, can capture important features
of information retrieval from the entangled-state model.

Recently, the {\it no-hiding theorem} \cite{me,Kretschmann} was used to
prove that Hawking's prediction of featureless radiation implied that the
information about the in-fallen matter could not be in the radiation
field, but must reside in the remainder of Hilbert space --- then
presumed to be the black hole interior. That work presented a strong
form of the black hole information paradox pitting the predictions of
general relativity against those of quantum mechanics \cite{me}. Here
we have shown that trans-event horizon entanglement provides a way out,
since now the ``remainder of Hilbert space'' comprises both the black
hole interior and external neighborhood modes. Because the evaporating
black hole actually involves three subsystems, the information may be
encoded within them as pure correlations via a quantum one-time
pad \cite{me,Leung02}: the information is in principle retrievable from
any two of the three subsystems, yet inaccessible from any single
subsystem alone.
This simultaneous encoding of information externally (in the combined
radiation and external neighborhood modes) and `internally' (if one
{\it stretches\/} the horizon to envelope the bulk of the external
neighborhood modes in addition to the black hole interior) is
reminiscent of Susskind's principle of black hole
complementarity \cite{Susskind93}. Yet, the overlap between the interior
and exterior in this picture eliminates any need for a temporary or
unobservable violation of the no-cloning theorem. Within the one-time
pad encoding trans-event horizon entanglement provides a mechanism
whereby Hawking's calculations may accurately describe the behavior
of out-going radiation from a black hole until very late in its
evaporation; it does not necessarily solve the paradox, but it delays
for as long as possible the clash between two of our most cherished
and fundamental theories of nature.

\vskip 0.1truein
\noindent
SLB gratefully acknowledges the hospitality of the Abe Watssman Institute
for Innovative Thinking, where this work was initiated. HJS and KZ
acknowledge financial support by the Deutsche Forschungsgemeinschaft under
the project SFB/TR12 and grant number DFG-SFB/38/2007 of the Polish
Ministry of Science and Higher Education.

\appendix

\section{Appendix A}
\label{decoupling}

We now summarize the decoupling theorem \cite{Abey06}.
Consider a pure state tripartite
$|\Psi\rangle_{\text{ref},\text{ext},A_1A_2}$,
where the joint subsystems $A_1A_2$ will be decomposed as either
the radiation modes and interior black holes modes $RB$ or vice-versa
$BR$. Tracing out the external neighborhood modes gives
\begin{equation}
\rho_{\text{ref},A_1A_2}\equiv \text{tr}_{\text{ext}}
\bigl(|\Psi\rangle_{\text{ref},\text{ext},A_1A_2} \langle\Psi|\bigr).
\end{equation}
Next, consider a (random) unitary applied to the joint subsystems $A_1A_2$.
This allows us to define
\begin{equation}
\sigma_{\text{ref},A_2}^U\equiv 
\text{tr}_{A_1} \bigl(U_{A_1A_2}\;
\rho_{\text{ref},A_1A_2}\; U_{A_1A_2}^\dagger\bigr).
\end{equation}
Then the decoupling theorem \cite{Abey06} states that
\begin{eqnarray}
&&\biggl(\int_{U\in U(A_1A_2)}dU\,\bigl\| \sigma_{\text{ref},A_2}^U-
\sigma_{\text{ref}}^U\otimes \sigma_{A_2}^U\bigr\|_1\biggr)^2\nonumber\\
&\le& \frac{A_2 K}{A_1}\bigl(\text{tr}\; \rho_{\text{ref},A_1A_2}^2
+\text{tr}\; \rho_{\text{ref}}^2 \;\text{tr}\; \rho_{A_1A_2}^2\bigr),
\end{eqnarray}
where states with `missing' subscripts denote further tracing out of
the relevant subspaces. Now
$1-F(\rho,\sigma) \le \frac{1}{2}\|\rho-\sigma\|_1$, where the trace
norm is defined by $\|X\|_1\equiv \text{tr}\, |X|$ and the fidelity
by $F(\rho,\sigma)\equiv \|\sqrt{\rho}\sqrt{\sigma}\|_1$.
As a consequence, the fidelity with which the entangled state
describing the in-fallen matter is encoded within the combined
$\text{ref},A_1,\text{ext}$ subsystem is bounded below by \cite{me}
\begin{equation}
\!1-\Bigl(\frac{A_2 K}{A_1}\, \text{tr}\; \rho_{\text{ext}}^2
\Bigr)^{\frac{1}{2}}, \label{nothing}
\end{equation}
where we use the fact that 
$\text{tr}\; \rho_{\text{ext}}^2=
\text{tr}\; \rho_{\text{ref},A_1A_2}^2\ge
\text{tr}\; \rho_{\text{ref}}^2 \;\text{tr}\; \rho_{A_1A_2}^2$
for our model. Eq.~(\ref{nothing}) expresses nothing more than the
lower bound to the fidelity with which the quantum state about the
in-fallen matter may be theoretically reconstructed from this joint
subspace. We note that the results for the pure-state model of black
hole evaporation \cite{Hayden07} may be recovered by setting
$\text{tr}\; \rho_{\text{ext}}^2=1$.

\section{Appendix B}
\label{heuristic}

The rigorous results from the manuscript may be heuristically visualized
by following how the correlations with the distant reference system
behave. For a pure tripartite state $XYZ$, these correlations satisfy
\begin{equation}
C(X\!:\!Y)+C(X\!:\!Z) = S(X), \label{monogamy}
\end{equation}
Here $S(X)$ is the von Neumann entropy for subsystem $X$ and
$C(X\!:\!Y)\equiv\frac{1}{2}[S(X)+S(Y)-S(X,Y)]$, one-half the quantum
mutual information, is a measure of correlations between subsystems
$X$ and $Y$. Relation~(\ref{monogamy}) is additive for a pure
tripartite state, so the correlations with subsystem $X$ smoothly
move from subsystems $Y$ to $Z$ and vice-versa.

\begin{figure}[ht]
\centering
\begin{tabular}{cc}
(a)$~~~~~~~~~~~~~~~~~~~~~~~~~~~~~~~~~$ &
(b)$~~~~~~~~~~~~~~~~~~~~~~~~~~~~~~~~~$ \\
  \begin{psfrags}
    \psfrag{qubitsradiated}[l]{$\scriptstyle \text{qubits radiated}$}
    \psfrag{XKB}[c]{$\scriptstyle ~~~~~~C(\text{ref}:B)$}
    \includegraphics[scale=0.45]{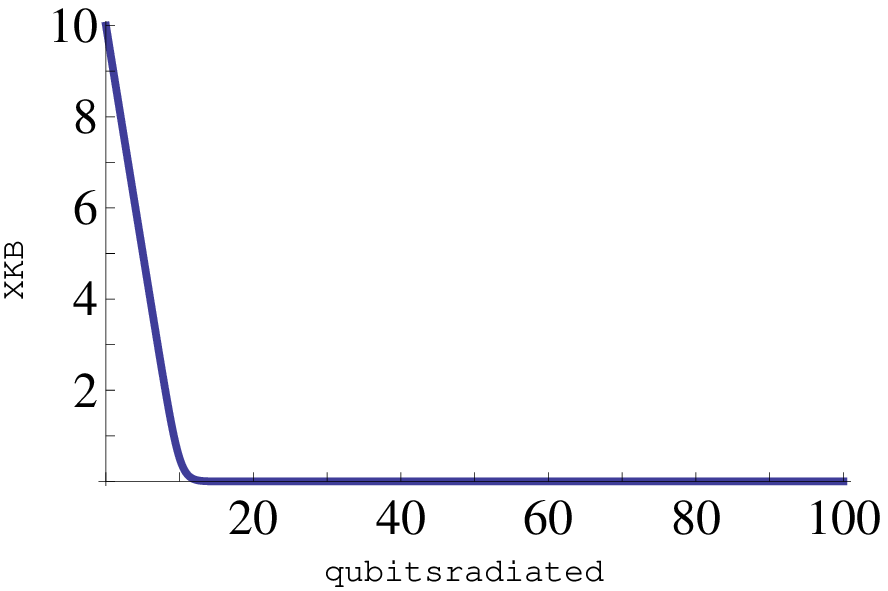}
  \end{psfrags} &
  \begin{psfrags}
    \psfrag{qubitsradiated}[l]{$\scriptstyle \text{qubits radiated}$}
    \psfrag{KAN}[c]{$\scriptstyle ~~~~~~C(\text{ref}:R,\,\text{ext})$}
    \includegraphics[scale=0.45]{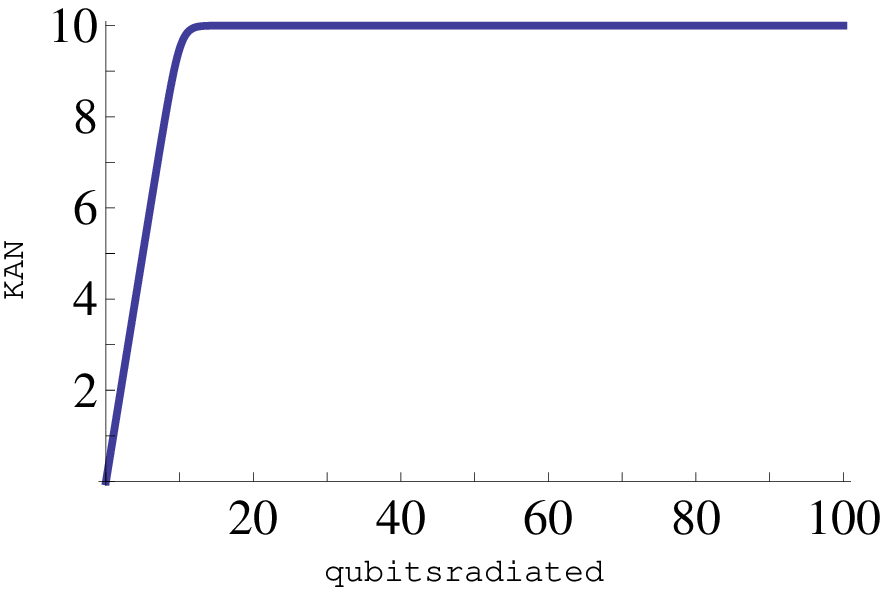}
  \end{psfrags} \\
(c)$~~~~~~~~~~~~~~~~~~~~~~~~~~~~~~~~~$ &
(d)$~~~~~~~~~~~~~~~~~~~~~~~~~~~~~~~~~$ \\
  \begin{psfrags}
    \psfrag{qubitsradiated}[l]{$\scriptstyle \text{qubits radiated}$}
    \psfrag{KBN}[c]{$\scriptstyle ~~~~~~C(\text{ref}:B,\,\text{ext})$}
    \includegraphics[scale=0.45]{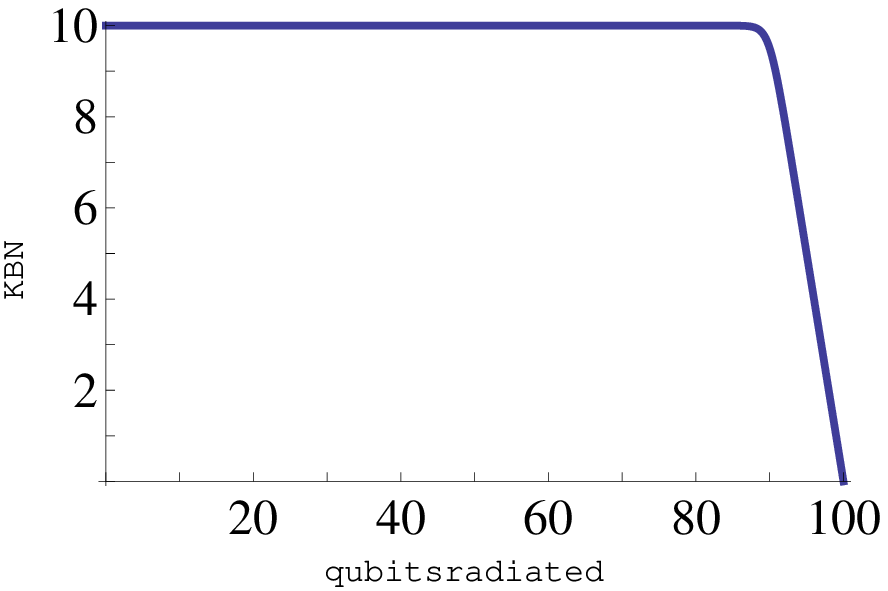}
  \end{psfrags} &
  \begin{psfrags}
    \psfrag{qubitsradiated}[l]{$\scriptstyle \text{qubits radiated}$}
    \psfrag{XKA}[c]{$\scriptstyle ~~~~~~C(\text{ref}:R)$}
    \includegraphics[scale=0.45]{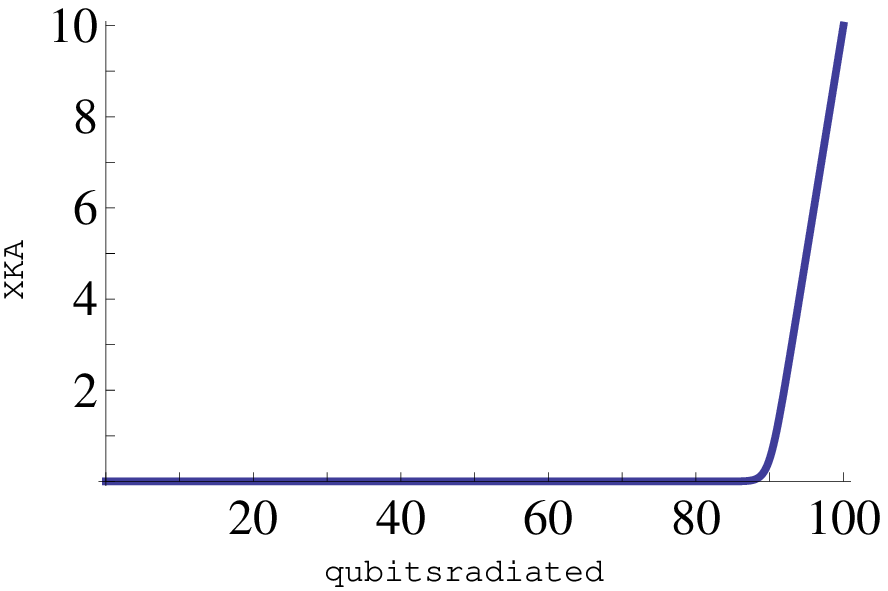} 
  \end{psfrags}
\end{tabular}
\caption{Correlations to the reference subsystem as a function of the
number of qubits radiated ($\log_2 R$). Correlations between the reference
(ref) subsystem and: (a) black hole interior, $B$; (b) radiation, $A$,
and external ($\text{ext}$) neighborhood modes; (c) black hole interior
and external neighborhood modes; and (d) radiation alone. Note that, as
expected from Eq.~(\ref{monogamy}), the sum of $C$'s in subplots (a)
and (b) is a constant, as is that of subplots (c) and (d). In each
subplot, the in-fallen matter consists of $k= 10$ qubits and
the black hole initially consists of $\log_2 RB = 100$ qubits
with $x=0$. (Entropies are evaluated using base-two logarithms.)}
\label{results}
\end{figure}

For simplicity, in this Appendix we restrict ourselves to the case
where
\begin{equation}
\rho_{\text{ext}}=\frac{1}{N}\sum_{j=1}^N
 |j\rangle_{\text{ext}}\,{}_{\text{ext}}\!\langle j|,
\end{equation}
and where $x=0$. We computed the above measure of correlations,
Eq.~(\ref{monogamy}), from von Neumann entropies approximated using the
average purity (see Appendix~\ref{purities} for details); numerical
calculations showed this as a good approximation for systems of even a
few qubits. Fig.~\ref{results} shows a typical scenario (assuming no
excess unentangled qubits): A black hole is assumed to be
created from in-fallen matter comprising $k$ qubits of information and
negligible excess unentangled qubits. Within the first $k$ qubits
radiated, information about the in-fallen matter (a) vanishes from the
black hole interior at roughly the radiation emission rate and (b) appears
in the joint radiation and external neighborhood subspace. From then
until just before the final $k$ qubits are radiated, the in-fallen
matter's information is encoded in a tripartite state, involving the
radiation, external neighborhood and interior subspaces, subplots~(b)
and~(c). In the final $k$ qubits radiated the information about the
in-fallen matter is released from its correlations and appears in the
radiation subsystem alone, subplot (d). This qualitative picture is
in excellent agreement with the results from the decoupling
theorem and its generalization.

\section{}
\label{purities}

In order to approximate the computation of the correlation measure
described in the text, we use a lower bound for a subsystem with
density matrix $\rho$
\begin{equation}
\langle\!\langle S(\rho) \rangle\!\rangle
\ge -\langle\!\langle \,\ln p(\rho) \rangle\!\rangle
\ge -\ln \langle\!\langle p(\rho) \rangle\!\rangle. 
\end{equation}
Here $S(\rho)=-{\text{tr}}\, \rho \ln \rho$ is the von Neumann entropy
of $\rho$, $p(\rho)= {\text{tr}}\, \rho^2$ is its purity, and here
$\langle\!\langle \cdots \rangle\!\rangle$ denotes averaging over
random unitaries with the Haar measure. The former inequality above is
a consequence of the fact that the R\'enyi entropy is a non-increasing
function of its argument \cite{Bengtsson06}, and the latter follows from the
concavity of the logarithm and Jensen's inequality. We may estimate
the von Neumann entropies required then by the rather crude
approximation
$\langle\!\langle S(\rho) \rangle\!\rangle
\approx -\ln \langle\!\langle \,p(\rho) \rangle\!\rangle$, 
which turns out to be quite reasonable for spaces with even a few qubits.

Although traditional methods \cite{Mello90} may be used to compute these
purities, a much simpler approach is to use the approach from
Ref.~\onlinecite{Abey06}. In particular, for a typical purity of interest
we use the following decomposition
\begin{eqnarray}
\text{tr}\; \sigma_{R,\text{ext}}^{U\;2}
&=& \text{tr} \bigl( \sigma_{R,\text{ext}}^U
\otimes \sigma_{R',\text{ext}'}^U \;
{\cal S}_{R,\text{ext};R',\text{ext}'}\bigr)\\
&=& \text{tr} \bigl( \rho_{\text{ref},RB,\text{ext}}\otimes
\rho_{\text{ref}',R'B',\text{ext}'}\nonumber\\
&&\times U_{RB}^\dagger\otimes U_{R'B'}^\dagger\,
{\cal S}_{R;R'}\, U_{RB}\otimes U_{R'B'}\,
{\cal S}_{\text{ext};\text{ext}'} \bigr) \nonumber
\end{eqnarray}
where ${\cal S}_{A;A'}$ is the swap operator between 
subsystems $A$ and $A'$, similarly,
${\cal S}_{AB;A'B'}={\cal S}_{A;A'}{\cal S}_{B;B'}$. Then the average
over the Haar measure is accomplished by an application of Schur's
lemma \cite{Abey06}
\begin{eqnarray}
&&\bigl\langle\!\bigl\langle U_A^\dagger\otimes U_{A'}^\dagger\;
{\cal S}_{A_2;A_2'}\; U_A \otimes U_{A'}\bigr\rangle\!\bigr\rangle
\nonumber \\
&=&\frac{A_2(A_1^2-1)}{A^2-1}\;\openone_{A;A'}
+\frac{A_1(A_2^2-1)}{A^2-1}\;{\cal S}_{A;A'}.
\end{eqnarray}
This approach allows us to straight-forwardly compute the required
purities as
\begin{eqnarray}
p(\text{ref})\!&=&\!\frac{1}{K},\quad
p(\text{ext})=\frac{1}{N},\quad
p(\text{ref,ext})=\frac{1}{KN},~~~~~ \nonumber \\
p(R)\!&=&\!\frac{1}{(RB)^2-1}
\Bigl( R(B^2-1)+\frac{B(R^2-1)}{KN}\Bigr), \\ 
p(R,\text{ext})\!&=&\!\frac{1}{(RB)^2-1}
\Bigl( \frac{R(B^2-1)}{N}+\frac{B(R^2-1)}{K}\Bigr), \nonumber
\end{eqnarray}
with $p(B,\text{ext})$ and $p(B,\text{ext})$ given by the
above expressions under the exchange $R\leftrightarrow B$, similarly
the exchange $K\leftrightarrow N$ gives us expressions for
$p(\text{ref},R)$, etc.

\end{document}